\documentclass[10pt]{article}
\usepackage{graphicx,floatflt,amssymb,epsfig,epsf} 
\def\be {\begin{equation}}
\def\ee {\end{equation}}
\def\bea {\begin{eqnarray}}
\def\eea {\end{eqnarray}}

\def\bra {\langle}
\def\ket {\rangle}
                                                                                
\def\g {\gamma}

\def\bbar {\overline{B}{}^0}

\def\bbbar {B^0-\bbar}

\def\bsss {b\to s\bar{s} s}

\def\bsbsbar {B_s-\overline{B_s}}

\def\pe {\perp}
\def\pa {\parallel}
\def\bbar {\overline{B}{}^0}

\def\ves {\varepsilon^\ast}
\def\lcfour {\varepsilon_{\mu\nu\alpha\beta}}
\def\L {\Lambda}
\def\Lzz {\Lambda_{00}}
\def\Lpez {\Lambda_{\pe0}}
\def\Lpaz {\Lambda_{\pa0}}
\def\Lpepe {\Lambda_{\pe\pe}}
\def\Lpapa {\Lambda_{\pa\pa}}
\def\Lpepa {\Lambda_{\pe\pa}}
\def\S {\Sigma}
\def\Szz {\S_{00}}
\def\Spez {\S_{\pe0}}
\def\Spaz {\S_{\pa0}}
\def\Spapa {\S_{\pa\pa}}
\def\Spepa {\S_{\pe\pa}}
\def\Spepe {\S_{\pe\pe}}

\def\Vzz {\rho_{00}}
\def\Vpez{\rho_{\pe0}}
\def\Vpaz{\rho_{\pa0}}
\def\Vpapa {\rho_{\pa\pa}}
\def\Vpepa{\rho_{\pe\pa}}
\def\Vpepe{\rho_{\pe\pe}}
\def\t {\theta}
\def\kst{K^\ast}
\def\msum {m_B + m_{\kst}}

\def\hnc {\frac{h_{NP}}{4N_c}e^{i\xi_{NP}}}

\def\opcit(#1){ {\em op. cit.}, #1}
\def\etal {\em et al.}

\def\issue(#1,#2,#3){#1 (#3) #2} 

\def\APP(#1,#2,#3){Acta Phys.\ Polon.\ \issue(#1,#2,#3)}
\def\ARNPS(#1,#2,#3){Ann.\ Rev.\ Nucl.\ Part.\ Sci.\ \issue(#1,#2,#3)}
\def\CPC(#1,#2,#3){Comp.\ Phys.\ Comm.\ \issue(#1,#2,#3)}
\def\CIP(#1,#2,#3){Comput.\ Phys.\ \issue(#1,#2,#3)}
\def\EPJC(#1,#2,#3){Eur.\ Phys.\ J.\ C\ \issue(#1,#2,#3)}
\def\EPJD(#1,#2,#3){Eur.\ Phys.\ J. Direct\ C\ \issue(#1,#2,#3)}
\def\EPL(#1,#2,#3){Europhys.\ Lett.\ \issue(#1,#2,#3)}
\def\IJMP(#1,#2,#3){Int.\ J.\ Mod.\ Phys. \issue(#1,#2,#3)}
\def\JHEP(#1,#2,#3){J.\ High Energy Physics \issue(#1,#2,#3)}
\def\JPG(#1,#2,#3){J.\ Phys.\ G \issue(#1,#2,#3)}
\def\MPL(#1,#2,#3){Mod.\ Phys.\ Lett.\ \issue(#1,#2,#3)}
\def\NP(#1,#2,#3){Nucl.\ Phys.\ \issue(#1,#2,#3)}
\def\NIM(#1,#2,#3){Nucl.\ Instrum.\ Meth.\ \issue(#1,#2,#3)}
\def\PL(#1,#2,#3){Phys.\ Lett.\ \issue(#1,#2,#3)}
\def\PRD(#1,#2,#3){Phys.\ Rev.\ D \issue(#1,#2,#3)}
\def\PRL(#1,#2,#3){Phys.\ Rev.\ Lett.\ \issue(#1,#2,#3)}
\def\SJNP(#1,#2,#3){Sov.\ J. Nucl.\ Phys.\ \issue(#1,#2,#3)}
\def\ZPC(#1,#2,#3){Zeit.\ Phys.\ C \issue(#1,#2,#3)}

\begin{document} 
\begin{flushright} 
CU-PHYSICS/16-2005\\
\end{flushright} 
\vskip 30pt 
 
\begin{center} 
{\Large \bf New Physics in \boldmath $b \to s \bar{s} s$\unboldmath~ Decay:\\
Study of \boldmath $B\to V_1 V_2$\unboldmath~ Modes}\\
\vspace*{1cm} 
\renewcommand{\thefootnote}{\fnsymbol{footnote}} 
{\large {\sf Soumitra Nandi} and {\sf Anirban Kundu}} 
 \\ 
\vspace{10pt} 
{\small 
   {\em Department of Physics, University of Calcutta\\
     92 Acharya Prafulla Chandra Road, Kolkata 700009, India}}\\  
 
\normalsize 
\end{center} 
 
\begin{abstract} 
Various nonleptonic decay channels mediated by the quark-level subprocess
$b\to s\bar{s} s$ show hints of deviation from the Standard Model expectations.
We analyse the double-vector decay $B\to\phi K^\ast$ with different generic
new physics structures and find the constraints on the parameter spaces
of new physics. The allowed parameter spaces are compatible with, but further
narrowed down from, those obtained
from a similar analysis using pseudoscalar modes. We also discuss further
predictions for this channel as well as for $B_s\to\phi\phi$, and show how even
a partial measurement of the observables may discriminate between different
models of new physics.
 
\vskip 5pt \noindent 
\texttt{PACS numbers:~ 13.25.Hw, 14.40.Nd, 12.15.Hh} \\ 
\texttt{Keywords:~~B Decays, B$_{\tt s}$ Meson, Physics Beyond Standard Model}
\end{abstract}

\renewcommand{\thesection}{\Roman{section}} 
\setcounter{footnote}{0} 
\renewcommand{\thefootnote}{\arabic{footnote}} 

\section{Introduction}

It has been felt for a long time that the data from the nonleptonic
decays of the $B$ meson, mediated by the transition $\bsss$, is not
what one expects from the Standard Model (SM) with the 
Cabibbo-Kobayashi-Maskawa (CKM) paradigm
for CP violation. On one hand this led to a vigorous exercise for the
understanding of low-energy QCD dynamics, including different models
for calculating the long-distance part of the amplitude, and on the
other hand this made such nonleptonic modes an ideal testing ground
for indirect signals of New Physics (NP) \cite{hiller}. 

Let us make it clear here and now that no data, taken alone, is a clear
indication for NP. Some of the discrepancies (like the value of $\sin(2\beta)$
extracted from $B\to\phi K_S$) are compatible with the SM predictions
at less than $2\sigma$, and some (like the abnormally large branching ratio
of $B\to\eta' K$) may be explained by SM dynamics yet to be fully understood.
However, all data, considered together, have a significant pull away from
the SM predictions, and one may hope that this will emerge as
an indirect signal of NP. This is more so since the $\bsss$ transition
involves quarks of the second and the third generations where the SM is less
well-tested. 

In an earlier publication \cite{kns}, we have shown that from a 
model-independent analysis of $B\to P_1P_2$ (two pseudoscalars) and
$B\to PV$ (one pseudoscalar and one vector) decays mediated by the
$\bsss$ transition, one can effectively constrain the parameter space
of NP, characterised by the strength of the NP coupling and its weak
phase. 

In this paper we focus upon the relevant decays of type $B\to V_1 V_2$,
which, in this case, are $B\to \phi K^\ast$ (with all charge combinations)
and $B_s\to\phi\phi$. These modes, in particular the former, have been
discussed in the context of specific NP models \cite{huang0511129} as well
as in a model-independent way, including possible modifications of
low-energy QCD dynamics \cite{phik-indep,yang}.
For $B\to\phi K^\ast$, data \cite{babar,hfag}
exists on branching ratios (BR),
different CP asymmetries, and different polarisation fractions (see Table
\ref{datahfag}). The error bars are still large but hopefully a much better
situation will arise in a few more years. For the $B_s$ decay we have only
some preliminary data \cite{hfag}
on its BR, but LHC-B should do a more thorough job.

The reason for such an analysis is twofold. First, one can construct more
observables than $B\to P_1P_2$ or $B\to PV$ cases, simply because the final
state mesons can be in $s$, $p$ or $d$-wave combinations. Since the 
wavefunctions have different parity, one can construct CP violating observables
even if the strong phase difference between various amplitudes be zero. 
Second, there are a few SM conditions \cite{lss} whose violations are
relatively simple to observe and which will indicate beyond any doubt the
presence of NP.

Experimentally, the anomalous trend persists in $B\to V_1V_2$ sector too. The
fraction of final states in a longitudinally polarised combination is
about 50\%, whereas one expects this to be dominant over the transverse
polarisation fractions, which are suppressed by the mass of the decaying
quark. At the infinite mass limit, all decays should be longitudinally
polarised.   

We must emphasise here that the longitudinal polarisation anomaly may
turn out entirely to be of SM origin. There are discussions in
the literature where contributions, neglected so far, have
been properly incorporated and their effects have been analysed. Needless
to say, most of the
insights have come {\em a posteriori}, after the experimental data is announced,
but that is only to be expected while dealing with something like low-energy
QCD. Kagan \cite{kagan0405134} has shown that the suppression of transverse
polarisation is still there in QCD Factorisation, but a new strong penguin
contribution can lower the longitudinal polarisation fraction in 
$B\to\phi K^\ast$. This, however, does not affect $B\to\rho K^\ast$, where
there is no such polarisation anomaly.  On the other
hand, Beneke {\em et al.} \cite{beneke0512258} have shown that there is 
a significant EW penguin contribution in $B\to\rho K^\ast$, and possibly 
the same mechanism works for $B\to\phi K^\ast$ too. 
Cheng {\em et al.} \cite{pqcd} have shown that in perturbative
QCD, the longitudinal polarisation can go down to 75\% if one takes annihilation
and nonfactorisable diagrams properly into account (without them it is
about 92\%). All in all, the explanation may lie within the SM, but there
is ample motivation to look for new physics.

Here we perform a model-independent analysis of the channel $B\to\phi K^\ast$
and extend the analysis to the SU(3)-related channel $B_s\to\phi\phi$. (Such
an analysis, with different set of operators, was also performed in \cite{yang},
and our conclusions are in agreement.) 
This is the first analysis of {\em all} anomalous $\bsss$ mediated decays in a 
model-independent way, alongwith predictions for $B_s\to\phi\phi$.
Analyses within the
framework of definite models (in particular different versions of SUSY) are
available, so is a partial model-independent analysis. The problem with
such partial analyses is that they give different allowed parameter space
for the NP for different anomalies. Of course, this does not mean that we
are evaluating the relevant amplitudes in a model-independent way!
                                                                                
In fact, this is the second part of the analysis. For the first part
involving $B\to P_1P_2$ and $B\to PV$ modes, we refer the reader to \cite{kns}.
We stress that we had to redo the analysis again since the data changed
in the last few months, in particular the $\sin(2\beta)$ anomaly. The results
have been summarised in Section IV, but we do not repeat the formalism which
obviously remains unchanged.

The
data on BR, CP asymmetries, and polarisation fractions are taken as input.
For the theoretical input, the major uncertainty occurs in the calculation
of long-distance contributions. We circumvent the problem by a rather
conservative approach. The NP effective Hamiltonian is characterised by a 
real positive coupling $h$, a NP weak phase $\xi$ (between 0 and $2\pi$),
and a Lorentz structure for the $\bsss$ current-current product. All 
short-distance corrections coming from the running to the NP scale to 
$m_b$ are dumped in $h$, but just for simplicity, we assume the NP
operator not to mix with the SM ones. The analysis not only gives the
allowed region in the $h$-$\xi$ plane, but also predicts the range
of different observables. Similar predictions are obtained for the
$B_s$ decay channel.


\section{SM and NP amplitudes}

The amplitude for $B(p)\to \phi(k_1,\varepsilon_1) + \kst(k_2,\varepsilon_2)$
can be written as \cite{amprefs,datta-london}
\be
{\cal M}=a\ves_1.\ves_2 + \frac{b}{m_B^2}(p.\ves_1)(p.\ves_2) + i\frac{c}
{m_B^2}\lcfour p^\mu q^\nu {\ves_1}^\alpha {\ves_2}^\beta
\label{amp-abc}
\ee
with $q=k_1-k_2$, and the CP-conjugate amplitude (for $\bar{B}$)
has the obvious form
\be
{\overline{\cal M}}=\bar a\ves_1.\ves_2 + \frac{\bar b}{m_B^2}(p.\ves_1)
(p.\ves_2) - i\frac{\bar c}
{m_B^2}\lcfour p^\mu q^\nu {\ves_1}^\alpha {\ves_2}^\beta,
\ee
where $a$, $b$, and $c$ are in general complex quantities, involving,
apart from the short-distance effects, the weak and the strong phases.

In the linear polarisation basis, we write
\be
{\cal M}({\overline{\cal M}})
=A_0 (\overline{A_0}) {\ves_1}^L {\ves_2}^L -\frac{1}{\sqrt{2}}A_{\pa}
(\overline{A_\pa})
\vec{{\ves_1}^T}.\vec{{\ves_2}^T} -(+)\frac{i}{2}A_\pe 
(\overline{A_\pe})\left(
\vec{{\ves_1}^T}\times \vec{{\ves_2}^T}\right).\hat{p}
\label{amp-lpol}
\ee
where $\hat{p}$ is the unit vector along $\kst$ in the rest frame of $\phi$, and
\be
{\ves_i}^L=\vec{\ves_i}.\hat{p},\ \
\vec{{\ves_i}^T}=\vec{\ves_i} - {\ves_i}^L \hat{p}.
\ee

The amplitudes of eq.\ (\ref{amp-lpol}) are related with those of eq.\
(\ref{amp-abc}) by
\be
A_\pa = \sqrt{2} a,\ \
A_0 = -ax - \frac{m_1m_2}{m_B^2}b(x^2-1),\ \
A_\pe = 2\sqrt{2} \frac{m_1m_2}{m_B^2} c\sqrt{x^2-1},
\ee
(and similarly for the barred variables) where
\be
x=\frac{k_1.k_2}{m_1m_2} = \frac{m_B^2-m_\phi^2-m_{\kst}^2}{2m_\phi m_\kst}.
\ee
In the SM, for the decay $B\to\phi\kst$, $a$ and $c$ are real and negative,
and $b$ is real and positive (assuming negligible strong phases), so that
$A_0>0$ (the $b$-term is mass-suppressed), while $A_\pa,A_\pe < 0$
\cite{datta-london}. Thus
one expects $\phi_\pe\equiv arg(A_\pe/A_0)\approx\pi$, $\phi_\pa\equiv
arg(A_\pa/A_0)\approx\pi$. However, these expectations may change due to
the presence of final state interactions (FSI).

An alternative formulation is in terms of the so-called helicity basis,
where the amplitudes are written in terms of $H_0$ and $H_\pm$, and
\be
H_0=A_0, \ \ \ H_{\pm} = \frac{A_\pa \pm A_\pe}{\sqrt{2}}.
\ee
The decay width is given by
\be
\Gamma =  \frac{|{\bf k}|}{8\pi m_B^2}
\left( |H_0|^2 + |H_+|^2 + |H_-|^2\right)
 =  \frac{|{\bf k}|}{8\pi m_B^2}
\left( |A_0|^2 + |A_\pa|^2 + |A_\pe|^2\right)
\ee
where {\bf k} is the magnitude of the three-momentum of either $V_1$ or $V_2$.
For the experimental observables, the amplitudes are normalised in such a way
that $|A_0|^2 + |A_\pa|^2 + |A_\pe|^2=1$, and similarly for the $\overline
{A_\lambda}$s \cite{lss} \footnote{This normalisation does not affect the
definition of the experimental observables in eq.\ (\ref{bellevar}).}.

To evaluate the transition amplitudes, we use the conventional factorisation 
(CF)
model \cite{bsw,ali}, with the standardised matrix elements as shown below
(with $q=k_1-k_2$ and $p=k_1+k_2$): 
\bea
\bra V_2|V_\mu | \bbar\ket &=& -\lcfour {\ves}^\nu p^\alpha k_2^\beta
\frac{2V}{m_B+m_{V_2}},\nonumber\\
\bra V_2|A_\mu | \bbar\ket &=& i\left(\ves_\mu-\frac{\ves.q}{q^2}q_\mu\right)
(m_B+m_{V_2})A_1 \nonumber\\
&{}& -i\left( (p+k_2)_\mu - \frac{m_B^2-m_{V_2}^2}{q^2}q_\mu\right) (\ves.q)
\frac{A_2}{m_B+m_{V_2}},\nonumber\\
\bra V_1|\bar{q}\g_\mu q|0\ket &=& f_{V_1} m_{V_1} \ves_\mu,\nonumber\\
\bra V_1|\bar{q}\sigma_{\mu\nu} q|0\ket &=&-if_{V_1}^T \left(
\ves_\mu {k_1}_\nu - \ves_\nu {k_1}_\mu\right),\nonumber\\
\bra V_2|\bar{q}\sigma_{\mu\nu}k_1^\nu (1+\g_5) b|\bbar\ket &=&
2iT_1\lcfour {\ves_2}^\nu p^\alpha k_2^\beta \nonumber\\
&{}& + T_2\left\{ {\ves_2}_\mu (m_B^2-m_{V_2}^2) -
\left( \ves_2.p\right) (p+k_2)_\mu\right\}\nonumber\\
&{}& +T_3\left(\ves_2.p\right) \left\{ k_{1\mu} - \frac{q^2}{m_B^2-m_{V_2}^2}
\left(p+k_2\right)_\mu \right\}.
\label{ffac}
\eea
The form factors, calculated in the light-cone sum rule (LCSR) approach, are
taken from \cite{ball}. Their values are given in Section III.

This gives, in the SM, the transition amplitude as
\bea
{\cal M}(B^-\to \phi K^{\ast -})&=&
{\cal M}(\bbar\to \phi \overline{K^{\ast 0}})\nonumber\\
&=& i\frac{G_F}{\sqrt{2}} f_\phi m_\phi \{- (\epsilon_\phi.\epsilon_{K^\ast})
(m_B+m_{K^\ast}) A_1^{B\to K^\ast}(m_\phi^2) \nonumber\\
&{}& +(\epsilon_\phi.p)(\epsilon_{K^\ast}.p)\frac{2A_2^{B\to K^\ast}(m_\phi^2)}
{(m_B+m_{K^\ast})}
-i\epsilon_{\mu\nu\alpha\beta}\epsilon_\phi^\mu \epsilon_{K^\ast}^\nu p^\alpha
k_2^\beta \frac{2V^{B\to K^\ast}(m_\phi^2)}{(m_B+m_{K^\ast})} \}
\nonumber\\
&\times& V_{tb}V_{ts}^\ast \{ a_3+a_4+a_5-\frac{1}{2}(a_7+a_9+a_{10})\}
     \label{bphikst1}
\eea
where the symbols have their usual meaning \cite{ali}. However, one may
question the validity of the NF approach for this decay, and indeed 
calculations based on QCD factorisation \cite{qcdf} or perturbative QCD
\cite{pqcd} indicate a discrepancy in the predicted BR by about a factor
of 2 at the most ($8.71\times 10^{-6}$ in QCD factorisation vis-\`a-vis
$14.86^{+4.88}_{-3.36}\times 10^{-6}$ in perturbative QCD for the decay
$B^0\to \phi K^{\ast 0}$).
To account for this, we have allowed the SM amplitude to
vary by 40\% for a fixed $N_c=3$ (this is equivalent to a 96\% variation
in the BR). Also note that this mode, like $B\to\phi K$, is not $N_c$-stable,
and the final result may have some quantitative variation for a different
$N_c$ \footnote{In the analysis, we stick to $N_c=3$ since the form factors
are evaluated \cite{ball} for this value.}. 
Apart from this variation, all other effects that may change the
predicted BR have been taken into account by varying the amplitude.

One may ask whether it would have been prudent to take QCD factorisation or
perturbative QCD as the model for the SM dynamics and estimate NP effects.
We would like to point out that they are models to calculate nonleptonic
decay amplitudes, just as CF, and though there is theoretical justification
for taking a particular model for a particular type of decay, there is no
reason to believe that for the decays in question, one is definitely
better than the others. So we stick to CF; however, it is necessary to take
into account the differences among various models, at least in a rough way.
This will provide for a larger uncertainty in the SM prediction. If we still
need NP to explain the data, we may be hopeful about its presence. The
question is how one does this. A good indicator is the predictions for the
BRs. We may hope that, roughly, the differences in BRs are reflected
in the amplitudes in different models. One may like to have some more 
error margin, since it is {\em better to be conservative
than over-ambitious}. The same is true for experimental data; in fact, we will
soon show that if the data is taken at $68\%$ confidence limit ($1\sigma$),
no possible NP can explain all the anomalies.

For the decay $B_s (p) \to\phi (k_1,\varepsilon_1) \phi
(k_2,\varepsilon_2)$, the transition amplitude is
\bea
{\cal M}(B_s\to \phi \phi)&=&
i\frac{G_F}{\sqrt{2}} f_\phi m_\phi
\{ -2(\epsilon_1.\epsilon_2) (m_{B_s}+m_{\phi})
A_1^{B_s\to \phi}(m_\phi^2) \nonumber\\
&{}& +(\epsilon_1.p)(\epsilon_2.p)
\frac{4A_2^{B_s\to \phi}(m_\phi^2)} {(m_{B_s}+m_{\phi})}
-i\epsilon_{\mu\nu\alpha\beta}\epsilon_1^\mu \epsilon_2^\nu
p^\alpha k_2^\beta \frac{2V^{B_s\to \phi}(m_\phi^2)}{(m_{B_s}+m_{\phi})} \}
\nonumber\\
&\times& V_{tb}V_{ts}^\ast \{ a_3+a_4+a_5-\frac{1}{2}(a_7+a_9+a_{10})\}.
   \label{bsphiphi1}
\eea
This is analogous to eq.\ (\ref{bphikst1}), with an extra factor of 2
for identical particles in the final state.

The amplitudes will have contributions from SM as well as from NP. Let us
write
\bea
&{}& A_\pa = R_1 e^{i\theta_1}, A_0=R_2  e^{i\theta_2}, A_\pe=R_3e^{i\theta_3},
\nonumber\\
&{}& \bar A_\pa= R_4 e^{i\theta_4}, \bar A_0=R_5e^{i\theta_5}, \bar
A_\pe=R_6e^{i\theta_6},
\eea
where $R_i$s and $\theta_i$s include all SM and NP effects (couplings, weak
and strong phases). The 18 variables proposed by \cite{lss} can be written
as
\bea
&& \Lzz=\frac{1}{2}\left(R_2^2+R_5^2\right),\ \ 
 \Lpapa=\frac{1}{2}\left(R_1^2+R_4^2\right),\ \ 
 \Lpepe=\frac{1}{2}\left(R_3^2+R_6^2\right),\nonumber \\ 
&& \L_{\pe 0}=R_2R_3\sin(\t_2-\t_3) - R_5R_6\sin(\t_5-\t_6),\nonumber \\ 
&& \L_{\pe\pa}=R_1R_3\sin(\t_1-\t_3) - R_4R_6\sin(\t_4-\t_6),\nonumber \\ 
&& \L_{\pa 0}=R_1R_2\cos(\t_1-\t_2) + R_4R_5\cos(\t_4-\t_5),\nonumber \\ 
&& \Szz=\frac{1}{2}\left(R_2^2-R_5^2\right),\ \ 
 \Spapa=\frac{1}{2}\left(R_1^2-R_4^2\right),\ \ 
 \Spepe=\frac{1}{2}\left(R_3^2-R_6^2\right),\nonumber \\ 
&& \S_{\pe 0}=R_2R_3\sin(\t_2-\t_3) + R_5R_6\sin(\t_5-\t_6),\nonumber \\ 
&& \S_{\pe\pa}=R_1R_3\sin(\t_1-\t_3) + R_4R_6\sin(\t_4-\t_6),\nonumber \\ 
&& \S_{\pa 0}=R_1R_2\cos(\t_1-\t_2) - R_4R_5\cos(\t_4-\t_5),\nonumber \\ 
&& \rho_{00} = R_2R_5 \sin(2\beta+\t_2-\t_5),\ \ 
 \rho_{\pa\pa} = R_1R_4 \sin(2\beta+\t_1-\t_4),\ \
 \rho_{\pe\pe} = -R_3R_6 \sin(2\beta+\t_3-\t_6),\nonumber \\ 
&& \rho_{\pe 0} = R_3R_5 \cos(2\beta+\t_3-\t_5) 
 + R_2R_6 \cos(2\beta+\t_2-\t_6),\nonumber \\ 
&& \rho_{\pe\pa} = R_3R_4 \cos(2\beta+\t_3-\t_4) 
 + R_1R_6 \cos(2\beta+\t_1-\t_6),\nonumber \\ 
&& \rho_{\pa 0} = R_1R_5 \sin(2\beta+\t_1-\t_5) 
 + R_2R_4 \sin(2\beta+\t_2-\t_4).
    \label{lssvar}
\eea
Here $\beta=arg(V_{td}^\ast)$ is the SM weak phase coming in $\bbbar$
mixing. We assume no NP contribution in this mixing. For the $B_s$ system,
$\beta_s$ is close to zero in the SM. However, NP of the type $\bsss$ may
contribute to $\bsbsbar$ mixing. Even then, the contribution of NP in mixing,
which is in effect a contamination to $\beta_s$, can hardly be worth
considering. The reason is this.
Only a lower bound on the SM amplitude exists. The NP amplitude with 
such a weak coupling as obtained from the decay fit to $B\to P_1P_2$
or $B\to PV$ modes can never compete with the SM
amplitude. We find $\sin(2\beta_s)$, the effective phase from the
$\bsbsbar$ box, to be never greater than $0.1$. (Similarly, $b\to c\bar{c}s$
channels are hopeless to look for new physics.) 

On the other hand, BaBar and Belle collaborations express their data
in terms of eight independent variables over which a fit is performed.
Apart from $f_L\equiv \Lzz$ and $f_\pe\equiv\Lpepe$, they are
\bea
A^0_{CP} &=& \frac{f^B_L-f^{\bar B}_L}{f^B_L+f^{\bar B}_L}
=\frac{R_2^2-R_5^2}{R_2^2+R_5^2},\nonumber\\
A^\pe_{CP} &=& \frac{f^B_\pe-f^{\bar B}_\pe}{f^B_\pe+f^{\bar B}_\pe}
=\frac{R_3^2-R_6^2}{R_3^2+R_6^2},\nonumber\\
\phi_\pa &=& \frac{1}{2}\left( arg(A_\pa/A_0) + arg(\bar{A}_\pa/\bar{A}_0)
\right) = \frac{1}{2}\left( \t_1-\t_2+\t_4-\t_5\right),\nonumber\\
\phi_\pe &=& \frac{1}{2}\left( arg(A_\pe/A_0) + arg(\bar{A}_\pe/\bar{A}_0)
\right) = \frac{1}{2}\left( \t_3-\t_2+\t_6-\t_5\right),\nonumber\\
\Delta\phi_\pa &=& \frac{1}{2}\left( arg(A_\pa/A_0) - arg(\bar{A}_\pa/\bar{A}_0)
\right) = \frac{1}{2}\left( \t_1-\t_2-\t_4+\t_5\right),\nonumber\\
\Delta\phi_\pe &=& \frac{1}{2}\left( arg(A_\pe/A_0) - arg(\bar{A}_\pe/\bar{A}_0)
\right) = \frac{1}{2}\left( \t_3-\t_2-\t_6-\t_5\right),
\label{bellevar}
\eea
where we have used a convention opposite to that used by BaBar, Belle and
HFAG to define the first two and last two variables of eq.\ (\ref{bellevar}).
These are the constraints that will go as inputs in our analysis. Note that
the set $\{-\phi_\pa,\pi-\phi_\pe,-\Delta\phi_\pa,-\Delta\phi_\pe\}$ 
is identical as far as the angular analysis is concerned. If we entertain the
possibility of NP, there is no reason to keep our analysis confined to the
set with values nearest to the SM expectation.

As in \cite{kns}, 
we discuss three different types of effective four-Fermi interactions
coming from new physics:
\bea
{\rm 1. ~Scalar:} \ \ &{}&
{\cal L}_{new} = h_s e^{i\xi_s}
\left( \bar{s}_{\alpha} (c_1+c_2\g_5) s_\alpha\right)
\left( \bar{s}_{\beta} (c_3+c_4\g_5) b_\beta\right),\nonumber\\
{\rm 2.~ Vector:} \ \ &{}&
{\cal L}_{new} = h_v e^{i\xi_v}
\left( \bar{s}_{\alpha} \g^\mu(c_1+c_2\g_5) s_\alpha\right)
\left( \bar{s}_{\beta} \g_\mu (c_3+c_4\g_5) b_\beta\right),\nonumber\\
{\rm 3. ~Tensor:} \ \ &{}&
{\cal L}_{new} = h_t e^{i\xi_t}
\left( \bar{s}_{\alpha} \sigma^{\mu\nu}(c_1+c_2\g_5) s_\alpha\right)
\left( \bar{s}_{\beta} \sigma_{\mu\nu}(c_3+c_4\g_5) b_\beta\right).
\eea
Here $\alpha$ and $\beta$ are colour indices. The couplings $h_{s,v,t}$
are effective couplings (generically denoted as $h_{NP}$), 
of dimension $[M]^{-2}$, that one obtains by
integrating out the new physics fields. They are assumed to be real and
positive and the weak phase information is dumped in the quantities
$\xi_{s,v,t}$ (again, generically denoted as $\xi_{NP}$), 
which can vary in the range 0-$2\pi$.
Note that they are effective couplings at the {\em weak} scale,
which one may obtain by incorporating all RG effects to the high-scale
values of them.
The couplings $c_1$-$c_4$ can take any values between
$-1$ and 1; to keep the discussion simple, we will discuss only six
limiting cases:
\bea
(i) \ (S+P)\times (S+P)\  [(V+A)\times (V+A)] &:&
c_1=1,\ c_2=1,\ c_3=1,\ c_4=1;
\nonumber\\
(ii)\  (S+P)\times (S-P)\  [(V+A)\times (V-A)] &:&
c_1=1,\ c_2=1,\ c_3=1,\ c_4=-1;
\nonumber\\
(iii)\  (S-P)\times (S+P) \ [(V-A)\times (V+A)] &:&
c_1=1,\ c_2=-1,\ c_3=1,\ c_4=1;
\nonumber\\
(iv) \ (S-P)\times (S-P) \ [(V-A)\times (V-A)] &:&
c_1=1,\ c_2=-1,\ c_3=1,\ c_4=-1;
\nonumber\\
(v) \ (T+PT)\times (T+PT) &:&
c_1=1,\ c_2=1,\ c_3=1,\ c_4=1;
\nonumber\\
(vi) \ (T-PT)\times (T-PT) &:&
c_1=1,\ c_2=-1,\ c_3=1,\ c_4=-1;
\eea
This choice is preferred since the $1-(+)\g_5$ projects out the weak doublet
(singlet) quark field. For the doublet fields, to maintain gauge invariance,
one must have an SU(2) partner interaction, {\em e.g.}, $\bar{s}(1-\g_5)s$
must be accompanied by $\bar{c}(1-\g_5)c$. No such argument holds for the
singlet fields. In the above equation, $PT$ denotes a pseudotensor structure,
characterised by $\sigma_{\mu\nu}\gamma_5$. 

The tensor current was not considered in \cite{kns}. Neither this form nor
its Fierz-reordered form can contribute to $B\to\phi K$. Now that with the
latest data \cite{hfag} one does not imperatively need NP for the
$B\to\phi K$ sector, one may feel justified to include this structure as well.
Note that only $\sigma^{\mu\nu}(1+(-)\gamma_5)\otimes \sigma_{\mu\nu}(1+(-)
\gamma_5)$ structures are of any interest; the other two, after Fierz
reordering, do not generate any scalar or pseudoscalar currents and hence
can affect none of the $B\to P_1P_2$ or $B\to PV$ modes. In fact, the
four-quark current $(T+(-)PT)\times (T-(+)PT)$ vanishes, which can be
checked from
the identity $\sigma^{\mu\nu}\gamma_5=-(i/2)\epsilon^{\mu\nu\alpha\beta}
\sigma_{\alpha\beta}$, with $\epsilon^{0123}=-1$ (though the factorised
matrix element need not vanish). 

We have chosen the interaction in a singlet-singlet form under SU(3)$_c$.
The reason is simple: one can always make a Fierz transformation to the
local operator to get the octet-octet structure. Note that the 
forms $(S+(-)P)\times (S+(-)P)$ 
generate tensor currents under Fierz reordering.
Such currents were not important in \cite{kns} since at least one of the
final state mesons was a pseudoscalar. Here it will be important since both
the final state mesons are spin-1 objects, and as we will see, the tensor
currents play a crucial role in bringing down the longitudinal polarisation
fraction of $B\to\phi K^\ast$. Since no such tensor current is available for
$(S+(-)P)\times (S-(+)P)$ type operators, or the vector-axial vector 
operators, there is no lowering of the longitudinal polarisation fraction.

We have kept the strong phase difference between the SM and the NP
amplitudes a free parameter. The short-distance strong phase, coming from the
imaginary parts of the respective Wilson coefficients, are calculable but
small. The long-distance strong phase, coming mostly from final-state 
rescattering, is a priori not calculable, but since there are not too many
final states of identical quark configuration, the strong phase is expected
to be not too large. However, there should not be any correlation between
the strong phase in $B\to\phi K^\ast$ and the strong phases in $B\to \phi K$
or $B\to\eta^{(')}K^{(\ast)}$, the channels discussed in \cite{kns}, but
the strong phase of $B_s\to\phi\phi$ can be related to that of $B\to\phi K^\ast$
by SU(3) symmetry. We will assume the breaking of flavour SU(3) to be small
and take equal strong phases in both these cases. The results are not at
all sensitive to a precise equality. 

The $a$, $b$, and $c$ terms of the NP amplitudes (eq.\ (\ref{amp-abc}))
for the decay processes $B\to\phi K^\ast$ and $B_s\to\phi\phi$
take the following form:

$B\to\phi K^\ast$
(the contributions are same for neutral and charged channels):\\
(i) Scalar-pseudoscalar channel
\bea
a_{NP}&=&i\left[ -f_\phi m_\phi (c_1c_4-c_2c_3) (\msum) A_1
+ 2f_\phi^T c_1c_4(m_B^2-m_{\kst}^2) T_2\right] \hnc,\nonumber\\
\frac{b_{NP}}{m_B^2}&=& i\left[ f_\phi m_\phi (c_1c_4-c_2c_3)\frac{2A_2}{\msum}
-c_1c_4 f_\phi^T \left(4T_2+4T_3\frac{m_\phi^2}{m_B^2-m_{\kst}^2}\right)
\right] \hnc,\nonumber\\
\frac{c_{NP}}{m_B^2}&=& i\left[- f_\phi m_\phi (c_2c_4-c_1c_3)\frac{V}{\msum}
+2 c_1c_3 f_\phi^T T_1 \right] \hnc.
\eea
Note the $1/N_c$ suppression; this channel can only contribute to the decay
after reordering. Also, note that for $(S+(-)P)\times (S+(-)P)$ structures
only the terms with tensor form factors survive, which is obvious since they
do not generate any vector or axialvector currents. 

(ii) Vector-axial vector channel
\bea
a_{NP}&=&
f_\phi m_\phi \left[ (c_1c_4+c_2c_3) + 4N_c c_1c_4\right] (m_B+m_{\kst})
A_1 \hnc,\nonumber\\
\frac{b_{NP}}{m_B^2}&=&
-f_\phi m_\phi \left[ (c_1c_4+c_2c_3) + 4N_c c_1c_4\right] \frac{2A_2}
{(m_B+m_{\kst})} \hnc,\nonumber\\
\frac{c_{NP}}{m_B^2}&=&
-f_\phi m_\phi \left[ (c_2c_4+c_1c_3) + 4N_c c_1c_3\right] \frac{V}
{(m_B+m_{\kst})} \hnc.
\eea

(iii) Tensor-pseudotensor channel
\bea
a_{NP}&=&-i\left[2f_\phi^T c_1c_4(m_B^2-m_{\kst}^2) T_2\right]
\left(1+\frac{1}{2N_c}\right) h_{NP} e^{i\xi_{NP}},\nonumber\\
\frac{b_{NP}}{m_B^2}&=& i\left[c_1c_4 f_\phi^T \left(4T_2+4T_3\frac{m_\phi^2}
{m_B^2-m_{\kst}^2}\right) \right] 
\left(1+\frac{1}{2N_c}\right) h_{NP} e^{i\xi_{NP}},\nonumber\\
\frac{c_{NP}}{m_B^2}&=& -i\left[2 c_1c_3 f_\phi^T T_1 \right]
\left(1+\frac{1}{2N_c}\right) h_{NP} e^{i\xi_{NP}}.
\eea

The expressions for $B_s\to\phi\phi$ are analogous, with the obvious
replacements $B\to B_s$, $\kst\to\phi$, and an extra factor of 2; see,
for comparison, eqs.\ (\ref{bphikst1}) and (\ref{bsphiphi1}). We do
not tabulate them separately.

\section{Theoretical and Experimental Inputs}

The experimental data, taken from \cite{hfag}, is shown in Table \ref{datahfag}.
The numbers are quoted for $B\to\phi K^\ast$ (neutral mode) while the 
corresponding numbers for charged $B$ decay, wherever they exist, are
given in parenthesis.
The error margins are shown at $1\sigma$ confidence limit (CL), 
while for the analysis, we have taken a more
conservative approach and kept the error margins at $2\sigma$.
We do not use the numbers that are derived from the primary measurements
assuming the validity of the SM, mostly $\Lpapa$, $\L_{\pa 0}$, and various
$\Sigma$s.
Note that since we are interested only in the $\bsss$ transition,
no other decay modes (like $B\to\rho\rho$) have been taken into consideration.

While the BR($B\to\phi K^\ast$) is in the expected ballpark, BR($B_s\to
\phi\phi$) is smaller than expected. The amplitude for the latter is twice
that of the former (identical particles in the final state) times the
SU(3) breaking effects, which gives an enhancement of the BR of the
latter by roughly a factor of 5.5-6. But the number of events for 
$B_s\to\phi\phi$ is small; it is compatible with zero at 95\% CL!
For analysis, we take this particular data with a $3\sigma$ error bar on
the higher side instead of the usual $2\sigma$, just to be cautious over the
preliminary nature of the data.

\begin{table}[htbp]
\begin{center}
\begin{tabular}{||c|c||c|c||}
\hline
 Observable & Value & Observable & Value \\
\hline
Br($B\to\phi K^\ast$) & $(9.5\pm 0.9)\times 10^{-6}$ &
Br($B_s\to\phi \phi$) & $(14^{+8}_{-7}))\times 10^{-6}$ \\
& $((9.7\pm 1.5)\times 10^{-6})$ & & \\
$\Lzz$ & $0.48 \pm 0.04$  & $f_\pe = \Lpepe$ & $0.26\pm 0.04$ \\
       & $(0.50\pm 0.07)$ &                  & $(0.19\pm 0.08)$ \\
$\phi_\pa$ & $2.36^{+0.18}_{-0.16}$ & $\phi_\pe$ & $2.49\pm 0.18$ \\
       & $(2.10\pm 0.28)$ &                  & $(2.31\pm 0.31)$ \\
$A^0_{CP}$ & $-0.01 \pm 0.08$  &  $A^\pe_{CP}$ & $0.16\pm 0.15$ \\
$\Delta\phi_\pa$ & $-0.03\pm 0.18$ & $\Delta\phi_\pe$ & $-0.03\pm 0.18$ \\
\hline
\end{tabular}
\caption{Data on $B\to\phi K^\ast$ modes, from \cite{hfag}. Our convention
of defining CP asymmetries is opposite to that of HFAG, see text. We do
not show, for obvious reasons, those observables which are not directly measured
but estimated using the validity of the SM.}
   \label{datahfag}
\end{center}
\end{table}

Apart from the BRs and CP asymmetries, we also use the following results from
\cite{hfag}:
\begin{itemize}
\item
$\sin(2\beta)$ from charmonium modes: $0.685\pm 0.032$;
\item
$\sin(2\beta)$ from $B\to K_S\phi$ transitions: $0.47\pm 0.19$ (the results do
not show any qualitative change if we use the combined $\bsss$ result:
$0.50\pm 0.06$; however, the averaging is a bit naive \cite{hfag} and should
be used with caution);
\item
$43.8^\circ < \gamma < 73.5^\circ$ at 95\% CL \cite{utfit}; this is needed for
a reevaluation of the constraints on the allowed parameter space (APS) of new
physics as found in \cite{kns}.
\end{itemize}
The CKM elements $V_{ts}$ and $V_{tb}$ are taken from \cite{pdg04}, with only
the unitarity constraint imposed. 

The constituent quark masses, in GeV, are taken to be \cite{ali}
\be
m_u=m_d=0.2,\ \ m_s=0.5,\ \ m_c=1.5,\ \ m_b=4.88,
\ee
though the final result is totally 
insensitive to the precise values. The constituent quark masses are
independent of the renormalisation scale. The Wilson
coefficients, evaluated at the regularisation scale $\mu=m_b/2\approx 2.5$
GeV, are also taken
from \cite{ali}. The corresponding current quark masses, which appear in
the Dirac equation for quarks while evaluating the hadronic matrix elements,
are (in GeV)
\be
m_u=0.0042,\ m_d=0.0076,\ m_s=0.122,\ m_c=1.5,\ m_b=4.88
\ee
at $\mu=2.5$ GeV (see table I of the first paper of \cite{ali}). The 
light quark masses will obviously shift if we evaluate them at some other
scale, say 1 GeV, and they also depend on how we evaluate them ({\em e.g.},
lattice QCD or QCD sum rules).
The decay constant of $\phi$, through vector and tensor currents, are defined
as
\bea
\bra 0| \bar{s} \g_\mu s |\phi(p,\lambda)\ket &=& f_\phi m_\phi \epsilon_\mu
^{(\lambda)},\nonumber\\
\bra 0| \bar{s} \sigma_{\mu\nu} s| \phi(p,\lambda)\ket &=&
if^T_\phi \left( \epsilon_\mu^{(\lambda)}p_\nu - \epsilon_\nu^{(\lambda)}p_\mu
\right),
\eea
and their numerical values (in GeV) are \cite{ball}
\be
f_\phi = 0.231 \pm 0.004,\ \ \ f^T_\phi = 0.200 \pm 0.010.
\ee
The form factors, evaluated in the light-cone sum rule (LCSR)
approach, are \cite{ball}
\bea
B\to K^\ast:&{}& 
A_1=0.292, A_2=0.259, V=0.411, T_1=T_2=0.333, T_3= 0.202,\nonumber\\
B_s\to \phi:&{}&
A_1=0.313, A_2=0.234, V=0.434, T_1=T_2=0.349, T_3= 0.175.
\eea
These form factors include the full twist-2 and twist-3 and the leading 
order twist-4 contributions. These numbers are for $q^2=0$. For nonzero
$q^2$, they change by about 10\%. We take the $q^2=0$ values for our
numerical evaluation. They are evaluated for $N_c=3$.

\section{Results}

\begin{table}[htbp]
\begin{center}
\begin{tabular}{||c|c|c||}
\hline
$ Lorentz $ & $h_{NP}$ & $\xi_{NP}$  \\
$structure$ & $(\times 10^{-8})$ & $({}^\circ)$ \\
\hline
$(S-P)\times(S+P)$ &$0.0-3.0$& $180-360$ \\
                   &$(0.5-3.0)$& $(185-340)$ \\
\hline
$(S+P)\times(S+P)$ &$0.0-4.5$& $0-180$ \\
\hline
$(S-P)\times(S-P)$ &$0.0-4.5$& $180-360$ \\
\hline
$(S+P)\times(S-P)$ &$0.0-3.0$& $0-180$ \\
\hline
$(V-A)\times(V+A)$ &$0.0-0.8$& $180-360$ \\
\hline
$(V+A)\times(V+A)$ &$0.0-2.1$& $0-180$ \\
                   &$(0.25-0.7)$& $(30-140)$ \\
\hline
$(V-A)\times(V-A)$ &$0.0-0.6$& $180-360$ \\
\hline
$(V+A)\times(V-A)$ &$0.0-2.2$& $0-180$ \\
                   &$(0.5-0.8)$& $(40-130)$ \\
\hline
\end{tabular}
\caption{Allowed parameter space from the analysis of $B\to P_1P_2$ and
$B\to PV$ modes, mediated by $\bsss$ transition (upgrade of \cite{kns}
in view of the Summer 2005 data \cite{hfag}). All error bars are taken to be 
at $2\sigma$. With $1\sigma$ error bars, only three structures survive,
as shown in parenthesis.}
    \label{knsredo}
\end{center}
\end{table}

Before we embark on an analysis of $B\to V_1V_2$ modes, let us revisit the
results of \cite{kns} in the light of Summer 2005 data. The main change is
the prediction of $\sin(2\beta)$ from $B\to\phi K_S$: this is now less than
$2\sigma$ away from the charmonium result. Thus, a nonzero NP amplitude based
on this data alone is no longer necessary (and hence tensor currents are
a possibility). Of course, the branching ratios of
the $\eta^{(')}K^{(\ast)}$ modes are still too large, but there is no way
one can reconcile that with a pure NP amplitude; there must be some dynamics
beyond the naive valence quark model \cite{kns}. (Lipkin \cite{lipkin}
has suggested that this is due to interferences between $B\to K\eta_8$ and
$B\to K\eta_1$ amplitudes, constructive for $\eta'$ and destructive for
$\eta$. Recently, an analysis based on the Soft Collinear Effective Theory
(SCET) claims this to be supported in that framework \cite{zupan}. However,
SCET is yet to provide numerical data on all the modes taken in our
analysis.) The allowed regions for
different Lorentz structures is shown in Table 2. Note that $h_{NP}$ can be
vanishingly small. However, the upper limits, which are controlled by the
BRs, remain unaltered. There does not seem to be any pressing need to introduce
new physics from this data alone, but we will soon find that the longitudinal
polarisation anomaly forces us to consider the NP option seriously. 

\begin{table}[htbp]
\begin{center}
\begin{tabular}{||c|c|c|c|c||}
\hline
Set  &  Lorentz  & $h_{NP}$ & $\xi_{NP}$ & $\delta_{NP}$  \\
No.\ & structure & $(\times 10^{-8})$& $({}^\circ)$ & $({}^\circ)$ \\
\hline
I & $(S+P)\times(S+P)$ &2.0 - 3.6& 0 - 30 &90 - 145\\
  & &             &  45 - 65  & 25 - 35   \\
  & &             & 150 - 180 & 270 - 325 \\
\hline
II & $(S-P)\times(S-P)$ &2.8 - 3.5& 271 - 278 &216 - 223\\
  & &             &  338 - 360  & 307 - 320   \\
\hline
III & $(T+PT)\times(T+PT)$ &0.14 - 0.36& 0 - 25 &140 - 170\\
    &                      &           & {\rm --- do ---}  &270 - 320\\
  & &             & 150 - 180 & 90 - 140 \\
    &                      &           & {\rm --- do ---} &320 - 350\\
    &                      &           & 45 - 70 &205 - 215\\
\hline
IV & $(T-PT)\times(T-PT)$ &0.20 - 0.27& 0 - 15 & 125 - 135 \\
    &                      &           &90 - 100 &215 - 225\\
    &                      &           &158 - 180 &305 - 320\\
\hline
\end{tabular}
\caption{Allowed parameter space for different Lorentz structures of
new physics. }
  \label{allowedset}
\end{center}
\end{table}

In Table \ref{knsredo} we show the principal allowed parameter space 
(APS) for different Lorentz
structures. For each structure, there is a subdominant
parameter space with very small
$h_{NP}$, but $\xi_{NP}$ in the opposite half-plane ({\em i.e.}, $\xi_{NP}
+\pi$ modulo $2\pi$). For example, for the structure $(S+P)\times (S+P)$
there is an APS with very small $h_{NP} \sim 10^{-9}$ and $\pi < \xi_{NP}
< 2\pi$. This was absent in \cite{kns}, but now that $\sin
(2\beta)$ from $B\to\phi K_S$ at $2\sigma$ may overshoot the charmonium
value, there is a scope for opposite interference. This is a general
trend for all structures.
However, these regions do not survive the $B\to V_1V_2$ analysis.
Also note that with $1\sigma$ error bars, three structures survive,
in contrast to only one as found in \cite{kns}.
Again, they disappear after the $B\to V_1V_2$ analysis.

With the expressions for BRs and CP asymmetries at hand, we perform a scan
on the new physics (NP) parameters $h_{NP}$ and $\xi_{NP}$. The starting
ranges for each Lorentz structure are shown in Table \ref{knsredo}. 
We perform a full scan over the tensor structures. This time,
however, we must introduce a nonzero strong phase difference between the SM
and the NP amplitudes (for simplicity, we assume this difference to be the 
same for all angular momentum channels). The reason is that both $\phi_\pa$
and $\phi_\pe$ differ from the SM expectation of $\pi$, even taking the
uncertainties $\Delta\phi$ into account. It is easy to see that if the strong
phase difference $\delta_{NP}$ is zero (modulo $\pi$), both $\phi_\pa$
and $\phi_\pe$ retain their SM expectation values. If the strong phase
is generated from rescattering, it should not be related in any way from
the strong phases in $B\to P_1P_2$ or $B\to PV$ channels; but from SU(3)
flavour symmetry, we expect the same $\delta_{NP}$ (at least to the leading
order) for both $B\to\phi K^\ast$ and $B_s\to\phi\phi$. In our analysis, we
take them to be the same. This effectively reduces our parameter set to
$h_{NP}$, $\xi_{NP}$ and $\delta_{NP}$.

Using the inputs discussed earlier, we find the APS for these three parameters
for different Lorentz structures. The result is shown in Table \ref{allowedset}.
Note that only two scalar-pseudoscalar and two tensor-pseudotensor channels
survive. This is due to the
fact that only these channels, under Fierz reordering, generate a tensor
current, which helps to bring down the longitudinal polarisation fraction 
$\Lzz$. This is in agreement with \cite{yang}.
These results are obtained with $2\sigma$ error bars; nothing
survives at $1\sigma$ CL. 

\begin{table}[htbp]
\begin{center}
\begin{tabular}{||c|c|c|c|c||}
\hline
Observable & Set I & Set II & Set III & Set IV \\
\hline
$\Lzz$ &  $0.40\to0.56$ & $0.40\to0.56$ & $0.40\to 0.56$ & $0.41\to 0.56$ \\
$\Lpepe$ &  $0.21\to0.29$ & $0.18\to0.26$& $0.21\to0.29$ & $0.18\to0.27$  \\
$\Lpapa$ & $0.23\to0.32$ & $0.21\to0.40$ & $0.23\to0.32$ & $0.23\to0.41$  \\
$\Lpez$ & $-0.15\to0.25$ & $-0.06\to 0.16 $ & $-0.25\to0.25$ & $-0.06\to0.18$ \\
$\Lpaz$ &  $0.25\to0.71$ & $0.46\to0.70$ & $0.25\to0.70$ & $0.46\to0.75$ \\
$\Lpepa$ &  $0.0$ & $-0.30\to 0.10$ & $0.0$ & $-0.29\to0.18$ \\
\hline
$\Szz$ &  $-0.07\to0.03$ & $-0.04 \to-0.005$ &$-0.07\to0.03$& $-0.04\to0.12$\\
$\Spepe$ &  $-0.01\to0.03$ & $0.01\to0.14$ &$-0.01\to0.03$  &$-0.03\to0.13$\\
$\Spapa$ &  $-0.01\to0.04$ & $-0.09\to-0.006$ &$-0.01\to0.04$ &$-0.09\to0.18$\\
$\Spez$ &  $0.22\to0.64$ & $0.44\to0.60$ & $0.24\to0.62$ &$0.37\to0.60$  \\
$\Spaz$ &  $-0.21\to0.22$ & $-0.37\to-0.05$ & $-0.20\to0.22$ & $-0.38\to0.12$ \\
$\Spepa$ &  $0.0$ & $0.005\to0.25 $ & $0.0$ & $0.005\to0.23$  \\
\hline
$\Vzz$ &  $-0.25 \to0.05$ & $-0.25\to0.05 $ &$-0.3\to0.05$  &$-0.26\to0.04$\\
$\Vpepe$ &  $-0.41\to-0.18$ & $-0.50\to-0.25$ & $-0.40\to -0.18$ & $-0.50\to-0.25$  \\
$\Vpapa$ &  $-0.27 \to0.05 $ & $-0.21 \to-0.04$&$-0.35 \to0.07$&$-0.35\to0.0$\\
$\Vpez$ &  $0.42\to0.70$ & $0.47\to0.67$ & $0.40\to0.70$ & $0.46\to0.67$ \\
$\Vpaz$ &  $-0.10\to0.42$ & $0.16\to0.27$ & $-0.10\to0.50$ & $0.14\to0.42$ \\
$\Vpepa$ &  $0.18\to0.70$ & $0.34\to0.60$ &$0.20\to0.65$  &$0.34\to0.58$  \\
\hline
\end{tabular}
\caption{Observables for $B\to\phi K^\ast$.}
\label{phikres}
\end{center}
\end{table}

\begin{table}[htbp]
\begin{center}
\begin{tabular}{||c|c|c|c|c||}
\hline
Observables& Set I & Set II & Set III & Set IV \\
\hline
$\Lzz$ &  $0.39\to0.66$ & $0.40\to0.60$ &$0.20\to0.65$  & $0.38\to0.58$ \\
$\Lpepe$ &  $0.14\to0.32$ & $0.19\to0.29$ &$0.15\to0.42$  & $0.18\to0.31$ \\
$\Lpapa$ &  $0.20\to0.33$ & $0.21\to0.36$ &$0.20\to0.38$  & $0.22\to0.36$ \\
$\Lpez$&$-0.70\to0.60$ & $-0.06 \to 0.16$&$-0.70\to0.60$ & $-0.09 \to 0.18 $ \\
$\Lpaz$ &  $0.10\to0.72$ & $0.52\to0.70$ & $0.10\to0.73$ & $0.49\to0.74$ \\
$\Lpepa$&$-0.08 \to0.07$&$-0.20\to 0.04$ &$-0.08 \to0.07 $  & $-0.20\to 0.06$ \\
\hline
$\Szz$ &$-0.026\to0.015$&$-0.04 \to-0.002$&$-0.030\to0.013$&$-0.04 \to0.008$ \\
$\Spepe$&$-0.025\to0.01 $ & $0.01 \to0.09 $&$-0.024\to0.01$&$-0.018\to0.09 $ \\
$\Spapa$&$-0.015\to0.04 $& $-0.06 \to-0.03$ &$-0.013\to0.04 $&$-0.06 \to0.01$\\
$\Spez$&$-0.24\to0.70$ &  $0.44\to0.66$ &$-0.22\to0.70$  & $0.36\to0.65$ \\
$\Spaz$&$-0.25\to0.25$ & $-0.37\to-0.04$ &$-0.30\to0.25$&$-0.38\to0.12$ \\
$\Spepa$&$-0.06\to0.08 $  & $0.004\to0.16 $ & $-0.07\to0.08 $&$0.004\to0.15$ \\
\hline
$\Vzz$ &  $-0.26\to0.25$ &$-0.13 \to 0.15$&$-0.26\to0.22$&$-0.15 \to 0.18$  \\
$\Vpepe$&  $-0.05 \to0.11 $&$-0.15\to 0.06$&$-0.04 \to0.11 $&$-0.15\to 0.05$  \\
$\Vpapa$&$-0.22\to0.30$ & $0.06 \to0.22$ &$-0.20\to0.24$  &$-0.14 \to0.25$  \\
$\Vpez$&$0.28\to0.62$ & $0.40\to0.75$ &$0.26\to0.74$  & $0.37\to0.77$ \\
$\Vpaz$&$-0.46 \to 0.25$ & $-0.40\to-0.22$ &$-0.47 \to 0.30$&$-0.42\to-0.14$  \\
$\Vpepa$&$-0.30\to0.50$ & $0.30\to0.55$ &$-0.30\to0.80$  & $0.28\to0.60$ \\
\hline
$A_{CP}^{\pe}$&$-0.15\to0.10$&$-0.22\to-0.02$&$-0.20\to0.10 $&$-0.23\to-0.02$ \\
$A_{CP}^{0}$&$-0.2 \to0.1  $ & $0.03\to0.25$ &$-0.20\to0.10 $  & $0.03\to0.25$\\
$\phi_{\pe}$&$-1.2 \to0.8  $ & $-0.95\to-0.82$ &$-1.2 \to0.8  $  &$-0.98\to-0.65$  \\
$\phi_{\pa}$&$-1.1 \to 0.6 $ & $-0.97\to-0.55$ &$-1.2 \to 0.6 $  &$-0.96\to-0.46$  \\
$\delta\phi_{\pe}$ &  $-1.5 \to 1.5$ & $-0.15 \to0.35 $ &$-1.3 \to 1.3$  & $-0.25 \to0.40 $ \\
$\delta\phi_{\pa}$ &  $-1.2 \to0.5 $ & $0.08\to0.36$ &$-1.2 \to0.4 $  &$-0.18\to0.3 $  \\
\hline
\end{tabular}
\caption{Observables for $B_s\to\phi\phi$.}
\label{phiphires}
\end{center}
\end{table}

In Tables \ref{phikres} and \ref{phiphires},
we show our main results: our expectations for the four allowed sets.
The results are quite similar for $B\to\phi K^\ast$ and $B_s\to\phi\phi$,
which is nothing but the manifestation of a rough SU(3) symmetry. In 
particular, we expect a similar suppression of $\Lzz$ for $B_s\to\phi\phi$
too.

At this point, one may like to look at the SM predictions for these observables.
In the conventional factorisation scheme, they are:\\
(i) $B\to\phi K^\ast$:
\bea
&{}& \Lzz=0.893,\ \Lpepe=0.051,\ \Lpapa=0.056,\ \Lpaz=0.446,\nonumber\\
&{}& \Vzz=-0.035,\ \Vpepe=-0.612,\ \Vpapa=-0.038,\nonumber\\
&{}& \Vpez=0.312,\ \Vpaz=0.306,\ \Vpepa=0.078.
\eea 
(ii) $B_s\to\phi\phi$:
\bea
&{}& \Lzz=0.948,\ \Lpepe=0.009,\ \Lpapa=0.044,\ \Lpaz=0.406,\nonumber\\
&{}& \Vpez=0.184,\  \Vpepa=0.039.
\eea 
All the other variables are zero. Again, this is in the conventional
factorisation model; nonfactorisable and annihilation contributions may reduce
$\Lzz$ of $B\to\phi K^\ast$ to about $0.75$ \cite{pqcd}. More variables are
zero for the latter case since we have two identical vector mesons and
the $B_s-\overline{B_s}$ mixing phase is close to zero in the SM.  

From tables \ref{phikres} and \ref{phiphires}, it appears that possible
four structures can be divided into two sets: one with $(S+P)\times
(S+P)$ and $(T+PT)\times (T+PT)$, and the other with $(S-P)\times
(S-P)$ and $(T-PT)\times (T-PT)$. 
Precise measurement of all the observables should be
able to discriminate between the two sets, but considering the respective
numbers, this is a more than formidable job. (An almost impossible
task is to discriminate between the pair of a given set. One way out is
to look for anomalies in semileptonic decays and analyse the angular 
distribution of the emitted leptons.)
On the other and, nonzero values
of most of these observables will point to new physics. Note that the values
of $\phi_\pa$ and $\phi_\pe$ are modulo $2\pi$, and the ambiguity of
$\{\phi_\pa,\phi_\pe,\Delta\phi_\pa,\Delta\phi_\pe\} \leftrightarrow
\{-\phi_\pa,\pi-\phi_\pe,-\Delta\phi_\pa,-\Delta\phi_\pe\}$
is still there.

This analysis makes some of the more favourite models of new physics
less so. A prime example is R-parity violating supersymmetry, which 
generates only $(S+(-)P)\times (S-(+)P)$ type interactions, but not
those that survive our analysis. 
The NP particles may be directly detected at the LHC if the corresponding
dimensionless couplings of the full theory are perturbative. For
example, if it is $\sim 0.1$, then we expect new particles to be
about 200-400 GeV, perfectly in the range of LHC. This is true even for
tensor currents if the tensor structure appears from an underlying 
radiative effect with loop suppressions coming into play.  

We would still like to mention again that the analysis should not be taken as
as irrefutable proof for and against certain types of new physics models.
We have tried to constrain the parameter space for generic NP models,
but unfortunately the SM uncertainty is still inordinately large. We have
taken a middle-of-the-way approach and tried to take into account the 
variations in predictions of different models (for amplitude calculation)
by incorporating some uncertainties in the SM prediction by hand. The
uncertainties are so chosen as to cover the BR predictions of different
models.
Hopefully, the estimates on NP parameter space
err on the conservative side. What one needs to
make the arguments watertight is to have more control on the theoretical
uncertainties. It is imperative that we have, as soon as possible, a 
particular theoretical model whose results can be relied upon upto a certain
order. This model should have a sound theoretical justification and 
{\em should not concentrate on reproducing the experimental results
alone} (in particular those that may contain hints of new physics). 
With this, and more data, can one hope to refine this analysis.

\section{Acknowledgements}

We thank K.-C. Yang for pointing out a mistake in the earlier version 
and for several useful discussions. 
A.K. thanks the Department of Science and Technology, Govt.\ of India,
for the project SR/S2/HEP-15/2003.


\begin{thebibliography}{99}

\bibitem{hiller}
See, {\em e.g.}, 
G. Buchalla {\etal}, \JHEP(09,074,2005).

\bibitem{kns}
A. Kundu, S. Nandi, and J.P. Saha, \PL(B622,102,2005).

\bibitem{huang0511129}
See, {\em e.g.}, C.-S. Huang {\etal}, hep-ph/0511129, for discussion 
of $B\to\phi K^\ast$ puzzle in MSSM, and also references therein.

\bibitem{phik-indep}
W.S. Hou and M. Nagashima, hep-ph/0408007;
H.n. Li, \PL(B622,63,2005);
C.H. Chen and C.Q. Geng, \PRD(71,115004,2005).

\bibitem{yang}
P.K. Das and K.C. Yang, \PRD(71,094002,2005).

\bibitem{babar}
B. Aubert {\etal} (BaBar Collaboration), hep-ex/0303020;
B. Aubert {\etal} (BaBar Collaboration), \PRL(91,171802,2003);
K.F. Chen {\etal} (Belle Collaboration), \PRL(91,201801,2003);
B. Aubert {\etal} (BaBar Collaboration), \PRL(93,231804,2004);
K.F. Chen {\etal} (Belle Collaboration), \PRL(94,221804,2005).


\bibitem{hfag}
See http://www.slac.stanford.edu/xorg/hfag/,
the website of the Heavy Flavour Averaging Group, for the summer 2005 update.

\bibitem{lss}
D. London, N. Sinha, and R. Sinha, \EPL(67,579,2004);
     \PRD(69,114013,2003).

\bibitem{kagan0405134} A.L. Kagan, \PL(B601,151,2004)

\bibitem{beneke0512258} M. Beneke, J. Rohrer, and D. Yang, hep-ph/0512258.

\bibitem{pqcd}
C.H. Chen, Y.Y. Keum and H.n. Li, \PRD(66,054013,2002).

\bibitem{amprefs}
G. Valencia, \PRD(39,3339,1989);
G. Kramer and W.F. Palmer, \PRD(45,193,1992),
     \PL(B279,181,1992),
     \PRD(46,2969,1992);
G. Kramer, W.F. Palmer, and T. Mannel, \ZPC(55,497,1992);
G. Kramer, W.F. Palmer, and H. Simma, \NP(B428,77,1994);
A.N. Kamal and C.W. Luo, \PL(B388,633,1996);
D. Atwood and A. Soni, \PRL(81,3324,1998),
    \PRD(59,013007,1999);
A.S. Dighe {\etal}, \PL(B369,144,1996);
B. Tseng and C.-W. Chiang, hep-ph/9905338;
N. Sinha and R. Sinha, \PRL(80,3706,1998);
C.-W. Chiang, \PRD(62,014017,2000);

\bibitem{datta-london}
A. Datta and D. London, \IJMP(A19,2505,2004).


\bibitem{bsw}
M. Wirbel, B. Stech, and M. Bauer, \ZPC(29,637,1985);
M. Bauer, B. Stech, and M. Wirbel, \ZPC(34,103,1987).

\bibitem{ali}
A. Ali, G. Kramer, and C.-D. L\"u, \PRD(58,094009,1998);
\PRD(59,014005,1999).

\bibitem{ball}
P. Ball and R. Zwicky, \PRD(71,014029,2005).

\bibitem{qcdf}
H.Y. Cheng and K.C. Yang, \PRD(64,074004,2001).

\bibitem{utfit}
See the UTfit website http://utfit.roma1.infn.it.

\bibitem{pdg04}
S. Eidelman {\etal} (Particle Data Group Collaboration), \PL(B592,1,2004).

\bibitem{lipkin}
H.J. Lipkin, \PL(B433,117,1998).

\bibitem{zupan}
A.R. Williamson and J. Zupan, hep-ph/0601214.

\end{thebibliography}
\end{document}